# Superconducting Diamond on Silicon Nitride for Device Applications

Henry A. Bland[1,*], Evan L. H. Thomas[1], Georgina M. Klemencic[1], Soumen Mandal[1], Andreas Papageorgiou[1], Tyrone G. Jones[2], Oliver A. Williams[1]

[1]School of Physics and Astronomy, Cardiff University, Queen's Building, The Parade, Cardiff CF24 3AA, United Kingdom.
[2]QMC Instruments Ltd., School of Physics and Astronomy, Cardiff University, Queen's Building, The Parade, Cardiff CF24 3AA, United Kingdom.
*Blandha@cardiff.ac.uk

Chemical vapour deposition (CVD) grown nanocrystalline diamond is an attractive material for the fabrication of devices. For some device architectures, optimisation of its growth on silicon nitride is essential. Here, the effects of three pre-growth surface treatments, often employed as cleaning methods of silicon nitride, were investigated. Such treatments provide control over the surface charge of the substrate through modification of the surface functionality, allowing for the optimisation of electrostatic diamond seeding densities. Zeta potential measurements and X-ray photoelectron spectroscopy (XPS) were used to analyse the silicon nitride surface following each treatment. Exposing silicon nitride to an oxygen plasma offered optimal surface conditions for the electrostatic self-assembly of a hydrogen-terminated diamond nanoparticle monolayer. The subsequent growth of boron-doped nanocrystalline diamond thin films on modified silicon nitride substrates under CVD conditions produced coalesced films for oxygen plasma and solvent treatments, whilst pin-holing of the diamond film was observed following RCA-1 treatment. The sharpest superconducting transition was observed for diamond grown on oxygen plasma treated silicon nitride, demonstrating it to be of the least structural disorder. Modifications to the substrate surface optimise the seeding and growth processes for the fabrication of diamond on silicon nitride devices.

**Introduction**

Silicon nitride is a well known technical ceramic of extreme hardness, wear resistance, resistance to thermal shock, strength, and fracture toughness, and has historically been used in the manufacture of cutting tools,[1] engine components,[2] and ball and roller bearings.[3] In theory, silicon nitride is an excellent substrate for the growth of CVD diamond due to their similar linear thermal expansion coefficients, which ensures the adhesion of diamond to the surface through a reduction in the interfacial stress.[4,5] As such, CVD diamond is regularly employed as a coating to further enhance the tribological properties of silicon nitride.[1] Recently however, progress has been made towards diamond on silicon nitride devices, employing the ceramic, for example, as an interlayer to facilitate polycrystalline diamond growth on GaN for GaN-based high electron mobility transistors.[6,7] The potential for integrated diamond – silicon nitride MEMS,[8-11] cryogenic radiation sensor arrays,[12] and graphene on silicon nitride transistors,[13] are also imagined. The fabrication of such device architectures would benefit from a comprehensive study of the silicon nitride surface, which could then be exploited, e.g. for the optimisation of CVD diamond nucleation and growth.

Heteroepitaxial diamond growth generally results in isolated diamond islands when a specific nucleation step is not undertaken.[14] Various techniques have thus been employed to enhance nucleation site density for diamond growth on silicon nitride in order to achieve homogeneous diamond thin films, with varying degrees of success. Mechanical abrasion[1,15-17] and ultrasonic micro-flawing,[18-21] using diamond particles, are amongst the most common nucleation techniques. Scratching of the substrate surface using diamond, in this manner, leaves behind small diamond fragments as nucleation sites,[22] and can produce nucleation densities up to $10^{11}$ cm$^{-2}$, required for the growth of diamond thin films.[23] However, by definition the above techniques lead to significant surface damage.[24] Whilst this is of little concern for high performance cutting tools, for thin film device applications, surface imperfections become wholly more significant, especially when the thermal and electrical properties of the diamond - silicon nitride interface are considered. The use of such seeding techniques is therefore inappropriate. Recent general nucleation efforts have focussed on the use of diamond seeds: nanoscale diamond particles synthesised through a detonation process, that are then dispersed over the substrate surface as nucleation sites. Polymer coatings have been utilised alongside diamond seeds to afford such a diamond rich growth surface,[25-28] although the technique raises questions over residual polymer contamination at the diamond – silicon nitride interface.

The self-assembly of a diamond seed monolayer onto a substrate through electrostatic forces of attraction is another common method of diamond seeding,[29,30] and extremely high seeding densities can be achieved on both 2D and 3D structures, if an understanding of the substrate surface is attained. Seeding densities for electrostatic seeding are modulated by the charging behaviour, or zeta potential ($\zeta$), of both the diamond seeds and the substrate surface, under aqueous conditions.[29] However, the surface charge of silicon nitride can vary considerably due to the reactive nature of silicon nitride under oxygen rich environments. Certainly, the Si-N bond is chemically reactive, and oxidation of the surface, although slow, is thermodynamically feasible at room temperature.[31] This paper therefore deals with the characterisation and optimisation of the silicon nitride surface ready for electrostatic seeding. We propose a pre-seeding treatment which allows for the homogenisation of the silicon nitride surface whilst also optimising the surface charge through modification of the surface functionality, allowing for the greatest possible seeding densities. In this work we correlate $\zeta$ potential and XPS analysis of the silicon nitride surface following exposure to an oxygen plasma, and we compare it to the surface following two standard cleaning techniques, the RCA-1 clean, and the solvent clean. The results are used to determine the best surface for the electrostatic self-assembly of a diamond nanoparticle monolayer. The seeded silicon nitride wafers are exposed to CVD growth conditions incorporating a gas-based boron dopant, and the resulting films are analysed under scanning electron microscopy (SEM), Raman spectroscopy, and their resistivity's are measured as a function of temperature, to allow for a comparison of each film's quality and structural properties.

**Results and Discussion**

The XPS survey spectra, and narrow scan spectra of the Si 2p peak, were taken for the silicon nitride surface following exposure to an oxygen plasma, a multi-stage solvent clean, and an RCA-1 cleaning solution, and are plotted in Figure 1. Table 1 details the atomic concentrations (at%) found at the silicon nitride surface following each treatment, determined by calculation of the area under the corresponding peaks in Figure 1a. The survey spectra present peak intensities for silicon (Si 2s, Si 2p), carbon (C 1s), nitrogen (N 1s), and oxygen (O 1s). The Si 2s and Si 2p peaks, at ~153 eV and ~102 eV respectively, remain approximately constant in intensity following each treatment. The C 1s peak at ~285 eV is also largely unchanged between the three treatments; carbon's relative concentration at the surface is extremely small, and can be explained by ambient contamination that occurs in most air stored samples.[32] The O 1s and N 1s peaks, at ~532 eV and ~397 eV respectively, show considerable variation in intensity between treatments. The binding energy of an element is modified by the electronegativity of its neighbouring atoms, and can lead to asymmetry in the photoelectron peak if more than one bonding environment is present. Deconvolution of the asymmetric Si 2p peak in Figure 1b therefore provides information on the chemical environment of the silicon atoms at the silicon nitride surface. Peak fitting in the Si 2p region reveals an intense peak at ~101.7 eV for each treated surface; labelled as Si-N, this peak is attributed to silicon nitride ($Si_3N_4$).[33,34] A secondary component of the Si 2p peak, labelled Si-O, at ~103.5 eV, is observed for silicon nitride following exposure to an oxygen plasma and solvent cleaning, and is assigned to $SiO_2$.[35] Oxygen plasma treated silicon nitride exhibits by far the larger of the two Si-O components, 26% of the area under the Si 2p peak, compared to 11% for that of the solvent cleaned surface. The RCA-1 cleaned surface does not display this peak at all within the resolution of the measurements taken.

The O 1s peaks in the XPS spectra of silicon nitride following exposure to an oxygen plasma and solvent cleaning, in Figure 1a, coupled with the asymmetry of the Si 2p peak in Figure 1b, is attributed to surface oxidation.[31] The more intense O 1s and Si-O peaks that appear following silicon nitride's exposure to an oxygen plasma would indicate a greater degree of surface oxidation, brought about by reactive oxygen species present in the plasma,[11,36] whilst the lesser oxidation of the solvent cleaned surface occurs passively under atmospheric conditions. Solvent cleaning is not expected to have an oxidising effect on the silicon nitride surface, and XPS analysis closely resembles that of silicon nitride wafers stored under atmosphere.[31] Atomic concentrations at the surface, presented in Table 1 suggest a relative increase in oxygen accompanies a relative decrease in nitrogen, indicating the replacement of silicon bound nitrogen with oxygen. Depth profiling of the partially oxidised silicon nitride surface, that which is comparable to our solvent cleaned surface, carried out by Holloway and Stein,[37] indicate the formation of a graded oxynitride film with a greater proportion of Si-O character at the surface, decreasing into bulk silicon nitride at greater wafer depths.[38]

The O 1s peak present in the XPS spectra of silicon nitride following RCA-1 cleaning is greatly decreased in intensity, to that on the order of contaminant carbon. An Si-O peak, corresponding to $SiO_2$, could not be fitted to the higher binding energy side of the Si 2p region, confirming a complete dissolution of the surface oxide. Kaigawa et al.[39] observed the ability of the RCA-1 cleaning solution to etch thermally grown $SiO_2$; ammonium hydroxide being determined to be active in etching $SiO_2$ a few years prior to the work.[40] The rate of $SiO_2$ etch by RCA-1 solution increased with both temperature and concentration of ammonium hydroxide in solution. Etch rates of ~0.25 nm/min were observed by the group, and would be more than sufficient to completely etch oxygen from the surface of our silicon nitride sample. Etching of the silicon nitride surface oxide has been observed in the literature, under treatment with hydrofluoric acid (HF), and resulted in a similar reduction in the O 1s line in XPS spectra, compared to the non etched surface.[31,38] The oxygen free surface is therefore ascribed to the etching ability of the RCA-1 cleaning solution, and we assign any residual oxygen at the surface to adsorbed oxygen species.

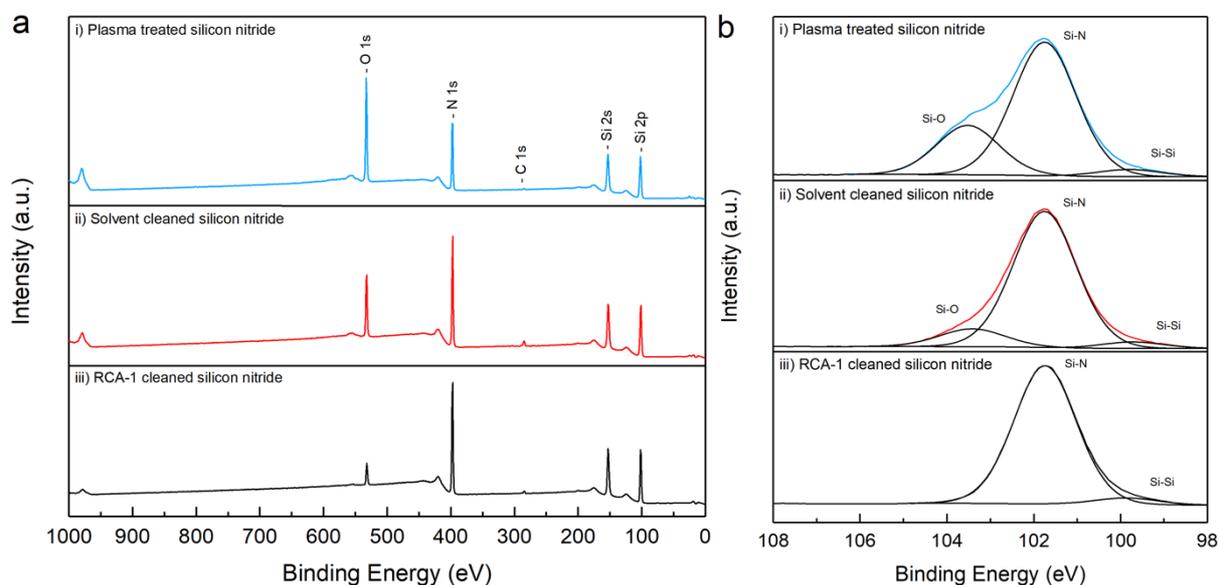

**Figure 1.** (a) XPS survey spectra of the silicon nitride surface following i) exposure to an oxygen plasma, ii) solvent cleaning, iii) RCA-1 cleaning. (b) XPS narrow scan spectra and chemical state analysis of the Si 2p region at ~102 eV, for the silicon nitride surface following each treatment.

|  | Oxygen Plasma | Solvent Clean | RCA-1 Clean |
|---|---|---|---|
| **Si 2p** | 43 | 44 | 47.5 |
| **N 1s** | 28 | 36 | 44.5 |
| **O 1s** | 28 | 16 | 6 |
| **C 1s** | 1 | 4 | 2 |

**Table 1.** Atomic composition (at%) of the silicon nitride surface following one of three pre-seeding treatments, determined from the integration of peaks in the XPS survey spectra of Figure 1a.

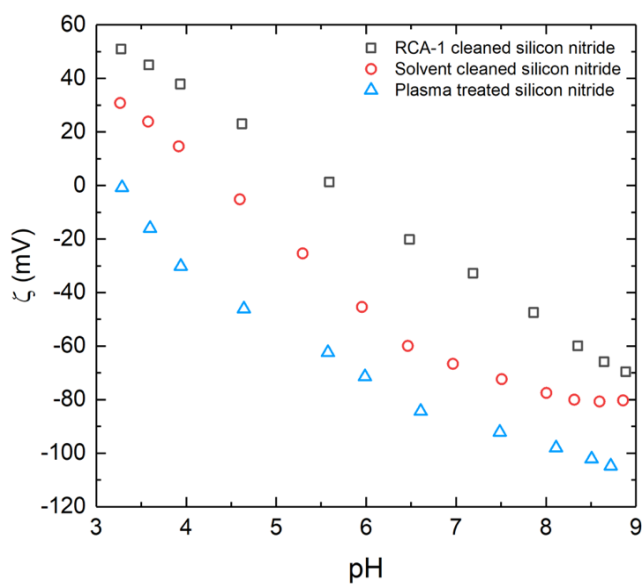

**Figure 2.** $\zeta$ potential measurements plotted as a function of pH, for the silicon nitride surface following exposure to an oxygen plasma, an RCA-1 cleaning solution, and solvent cleaning. Isoelectric points (pH$_{IEP}$) are extrapolated and interpolated from the data trends; they are found at pH ~3.2, ~5.7, and ~4.4, respectively.

The $\zeta$ potential is a measure of the overall charge exhibited by a surface or particle. Figure 2 details $\zeta$ potential measurements of the silicon nitride surface following each of the three surface treatment methods. A decrease in $\zeta$ potential is exhibited for all silicon nitride surfaces as a function of increasing pH, with a variation of ~100 mV between pH extremes. Following exposure to an oxygen plasma, silicon nitride displays the most negative $\zeta$ potential across the measured range, with an extrapolated pH$_{IEP}$, the point at which the surface exhibits zero net charge, of ~3.2; approaching that of SiO$_2$.[41] The solvent cleaned and RCA-1 cleaned surfaces see interpolated pH$_{IEP}$ = ~4.4 and pH$_{IEP}$ = ~5.7, respectively. Changes to the pH$_{IEP}$ may be attributed to the relative concentrations of functional groups present at the silicon nitride surface following treatment. Functional groups undergo acid/base type reactions under certain aqueous conditions to provide the surface with an overall charge. The oxygen plasma treated surface exhibits the lowest overall $\zeta$ potential, coinciding with the greatest degree of surface oxidation and SiO$_2$ character, determined from Figure 1. The surface is therefore likely to present functional groups similar to that of SiO$_2$, believed to be silanol (Si-OH) groups,[42,43] which will dissociate their hydrogen atom under certain aqueous conditions, to afford the surface with the measured negative charge. Chingombe et al.[44] and Garcia et al.[45] found that following chemical oxidation of the surface of activated carbon, $\zeta$ potential measurements observed a marked decrease towards negative potentials, when compared to measurements of an equivalent unoxidised surface. We therefore attribute the pH$_{IEP}$ of the solvent cleaned silicon nitride surface to a reduction in the concentration of surface silanol groups due to a lesser degree of surface oxidation. The etched surface of the RCA-1 cleaned silicon nitride displays a $pH_{IEP}$ of ~5.7, and would fit with our interpretation of the etched oxide free surface, and corresponding surface functionality. Bousse et al.[46] observed a similar increase in ζ potential after etching of the silicon nitride surface with hydrofluoric acid (HF) to remove the surface oxide.

For the self-assembly of a nanodiamond monolayer onto a growth substrate, it is important to achieve opposing charges and the greatest disparity in ζ potential between the substrate surface and the diamond seed solution. Whilst 5 nm hydrogen-terminated diamond particle solutions have been shown to exhibit a ζ potential of +40 mV,[29] in practice a value closer to +30 mV is observed. With water being the solvent of choice for the seed solution, the main region of interest in Figure 2 lies between pH 6 and 7. Following exposure to an oxygen plasma, the silicon nitride surface exhibits the most negative ζ potential in this region, with an averaged value for the region of -80 mV. After solvent cleaning, the silicon nitride surface exhibits an averaged ζ potential of -60 mV, and following RCA-1 cleaning, a value of -20 mV. Oxygen plasma treated silicon nitride therefore displays the greatest disparity in ζ potential, optimal for high density diamond seeding.

In order to establish the effectiveness of each pre-seeding treatment in promoting high seeding densities and uniform seed coverage, boron-doped diamond thin films were grown on silicon nitride following seeding with hydrogen-terminated diamond particle solution. Figure 3 displays scanning electron microscopy (SEM) images of boron-doped nanocrystalline diamond thin films grown following exposure to (a) an oxygen plasma (b) solvent cleaning (c) RCA-1 cleaning. Micrographs (a) and (b) show fully coalesced films with average grain sizes of 118 nm and 111 nm, respectively. The boron-doped diamond film grown on silicon nitride following RCA-1 cleaning exhibits pin-holing and a far greater average grain size of approximately 150 nm. Pin-holing and a larger average grain size are attributed to lower seeding densities following RCA-1 cleaning, stemming from weaker forces of electrostatic attraction between the silicon nitride surface and the diamond seeds, as indicated by ζ potential measurements in Figure 2. Of greater significance to granular superconductors such as boron-doped diamond, is the grain size distribution of each film. Plot (d) in Figure 3 displays the frequency with which various grain sizes are present in each film. The diamond film grown on RCA-1 cleaned silicon nitride exhibits by far the greater variation in grain size compared to films grown following exposure to oxygen plasma and solvent cleaning, and again this may be ascribed to lower seeding densities brought about by a far less negative substrate surface charge.

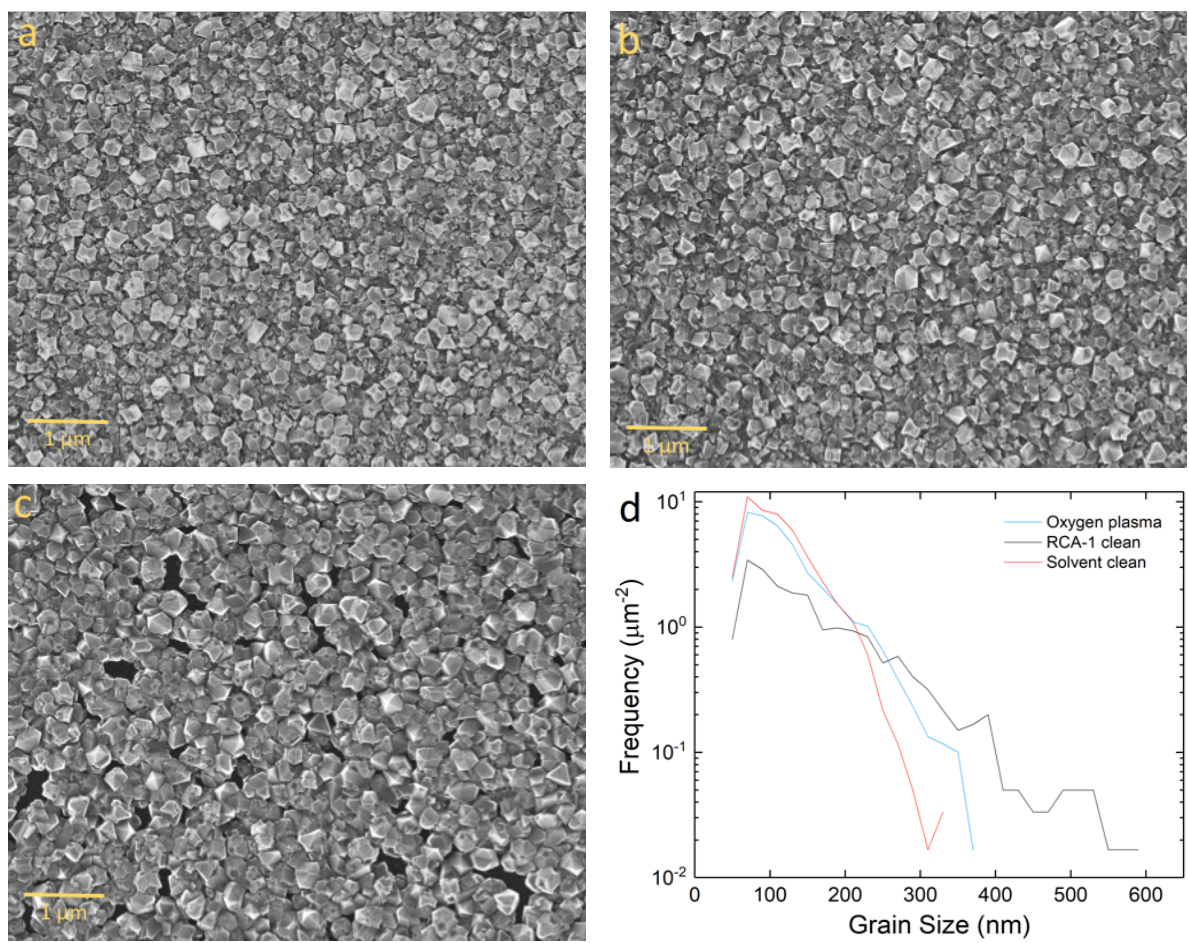

**Figure 3.** SEM images of boron-doped diamond grown on silicon nitride following (a) exposure to an oxygen plasma (b) solvent cleaning (c) RCA-1 cleaning. Clear pin-holing is exhibited for diamond grown on RCA-1 cleaned silicon nitride. Plot (d) displays the grain size distribution for each diamond film.

To assess the quality of the films produced, we employ Raman spectroscopy. Raman spectroscopy allows for the determination of the chemical structure of the boron-doped diamond films, after ~1 hour of CVD growth. Normalised to the resonance around 1220 cm$^{-1}$, the Raman spectra of heavily boron-doped nanocrystalline diamond films grown on silicon nitride following three pre-seeding treatments are offset and plotted in Figure 4. An excitation wavelength of 514 nm ensures the resonance of both sp$^2$ and sp$^3$ sites within the film.[47] As is to be expected for thin diamond films, the second order silicon peak of the underlying substrate, is visible at 980 cm$^{-1}$ due to the penetration depth of the Raman laser.[48] The first order silicon peak lies beyond the range of these measurements. A shoulder peak visible at around 950 cm$^{-1}$ is regularly observed for thin film polycrystalline diamond grown on silicon based substrates.[49,50] The feature at 850 cm$^{-1}$ is attributed to amorphous silicon carbide, SiC, with a similar broad peak being observed in the literature for silicon carbide films at approximately 820 cm$^{-1}$. A shift to higher wavenumbers can be justified by changes to the bonding state, manifesting as a greater degree of microcrystallinity.[51,52]

Typical of heavily boron-doped polycrystalline diamond are the broad band, B, centred around 1220 cm$^{-1}$, and the shoulder, D$_F$, at 1285 cm$^{-1}$.[53] Although typical, the origin of the broad band at 1220 cm$^{-1}$ is somewhat controversial, being attributed in the literature to both boron dimers and boron-carbon bonding states. Sidorov and Ekimov,[54] provide evidence to the contrary, following isotopic shift studies with boron isotopes, finding that the shift must be attributed to carbon-carbon bonding states. A correlation between band intensity and boron doping levels, however, has been observed,[50] and we therefore attribute the band to locally distorted lattice structures induced by the addition of the dopant boron.[55] The shoulder at 1285 cm$^{-1}$ is attributed to the crystalline diamond lattice. The diamond line is characteristically red shifted for boron-doped diamond, from 1332 cm$^{-1}$ to 1285 cm$^{-1}$, as well as appearing broader in nature. The diamond line adopts an asymmetric Fano-like line shape, due to the quantum interference between the triply degenerate zone centre diamond phonon and the continuum of electronic states induced by the presence of boron defects.[56,57] A further shoulder peak at 1120 cm$^{-1}$ is attributed to *trans*-polyacetylene, TPA, lying at the grain boundaries.[58]

The disordered carbon peak, D, at 1405 cm$^{-1}$ is ascribed to the breathing mode of graphitic rings, whilst the G peak at 1560 cm$^{-1}$ is the result of the in-plane bond stretching mode of a pair of sp$^2$ carbon sites.[59,60] The G band is shifted slightly from 1580 cm$^{-1}$ to 1560 cm$^{-1}$, due to the conversion of sp$^2$ sites to sp$^3$ sites that accompanies the transition of π ring systems to π chain systems.[61] The Raman spectra would suggest that diamond grown on oxygen plasma treated silicon nitride incorporates fewer sp$^2$ sites compared to that of other treatments, suggesting a more structurally ordered film. Certainly, when compared to the diamond peak at 1285 cm$^{-1}$, diamond grown on RCA-1 cleaned silicon nitride displays far more intensity in its D and G band structure, suggesting a greater degree of structural disorder in the film.

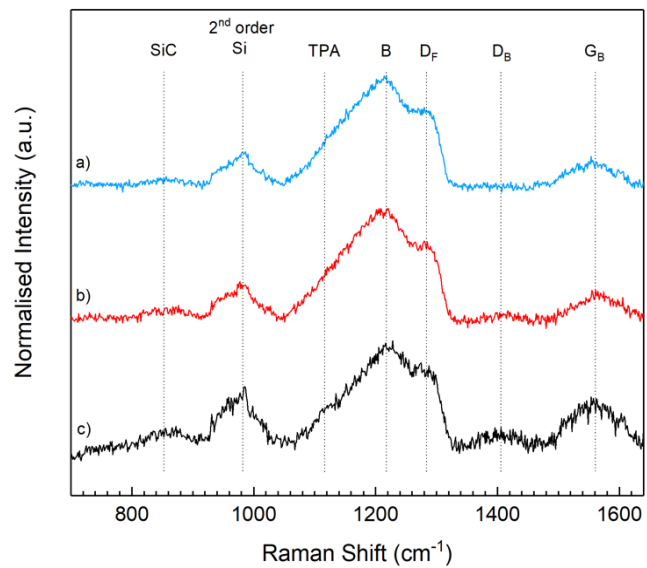

**Figure 4.** Raman spectra of boron-doped diamond grown on silicon nitride following exposure to (a) oxygen plasma treatment (b) solvent cleaning (c) RCA-1 cleaning, and diamond seeding. The poor signal to noise ratio is attributed to a high boron content and to the thickness of the diamond films.

For many superconductive device architectures, the resistive transition temperature, $T_C$, and the resistive transition width, $\Delta T_C$, are of fundamental importance. To assess the extent to which the three pre-seeding treatments affect these variables, the resistance of each film is measured as a function of temperature. Figure 5a displays a plot of the resistance against temperature in the temperature range 1.9 - 300 K. As the temperature is modulated, clear differences between the relative resistance of each film emerges, due to variations in intergranular connectivity. A comparatively lower normal state resistance, such as that exhibited by boron-doped diamond on oxygen plasma treated silicon nitride, can be attributed to a higher connectivity of the superconducting crystals of the polycrystalline diamond film. Furthermore, the sharp increase in resistance observed at approximately 50 K, for each film, can be attributed to the underlying silicon nitride substrate, and is not observed for boron-doped diamond on more common substrates, such as silicon.

A plot of the normalised resistance as a function of temperature is shown in Figure 5b. The inset in Figure 5b shows a plot of the first derivative of each resistance curve against temperature. The measured resistance decreases for each sample as the temperature is reduced from 7 K to 2.5 K. We determine the transition temperature to be the point at which the resistance diverges from zero.[62] Boron-doped diamond grown on oxygen plasma treated silicon nitride and on solvent cleaned silicon nitride exhibit equivalent transition temperatures, $T_C$ = 3.5 K, but vary in their transition widths, $\Delta T_C$ = 1.2 K and $\Delta T_C$ = 1.5 K, respectively. The peak height of the derivative of the oxygen plasma treated sample's resistance curve, shown in the inset of Figure 5b, is far higher than that of the solvent cleaned or RCA-1 cleaned films, and confirms a much sharper transition. Following the RCA-1 cleaning of silicon nitride, the diamond film exhibits a significantly reduced transition temperature compared to the other films, $T_C$ = 2.5 K, and an increased transition width, $\Delta T_C$ = 1.8 K. The wider transition width and lower transition temperature of the RCA-1 cleaned sample may be attributed to a greater degree of structural disorder within the film, compared to the other samples. This interpretation is in good agreement with grain size analysis from Figure 3c&d where we observed pin-holing and a greater variation in grain size for the RCA-1 cleaned sample, both of which contribute to the level of structural disorder. In contrast, the higher transition temperature and lower transition width of the plasma treated sample would indicate a far more ordered film.

Further to this, we observe a correlation between increasing $sp^2$ character in the Raman spectra of Figure 4 and the transition widths in Figure 5b, which we again link to an increase in the structural disorder of the respective film. Again, the most uniform film is produced following oxygen plasma treatment of the silicon nitride surface.

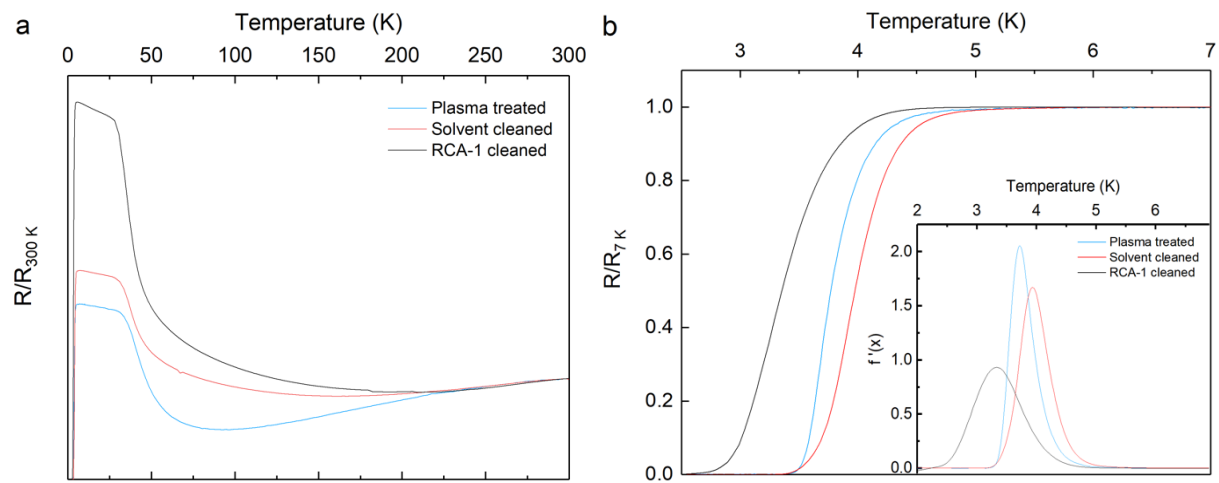

**Figure 5.** Resistance vs temperature plots for boron-doped diamond thin films grown on silicon nitride following three pre-seeding treatments (a) for temperatures between 1.9 K and 300 K, where each dataset is normalised to $R_{300\ K}$ (b) between 2.5 K and 7 K, where each dataset is normalised to $R_{7\ K}$. The first derivative of each resistance curve is plotted against temperature, in the inset.

## Conclusions

The surface of silicon nitride has been chemically modified to allow for higher seeding densities and fully coalesced films comprising a more uniform grain size. XPS and ζ potential measurements allowed for a full description of the surface modification and surface charge. Following exposure to an oxygen plasma, silicon nitride exhibits an oxidised surface, and an extremely negative surface charge; $pH_{IEP}$ = 3.2. Following solvent cleaning, the partial oxidation of the surface from ambient oxygen based species is observed with a corresponding $pH_{IEP}$ = 4.4. The silicon nitride surface oxide was found to be etched by RCA-1 cleaning solution, and ammonium hydroxide is proposed to be the active agent. The etched surface exhibits a $pH_{IEP}$ = 5.7. Boron-doped diamond films were grown following each pre-seeding treatment, and fully characterised under SEM, Raman spectroscopy, and their resistance measured as a function of temperature. Clear pin-holing was observed on samples that had undergone RCA-1 cleaning, whilst coalesced films were observed for solvent cleaning and oxygen plasma treatment. Resistivity data shows a clear improvement between RCA-1 cleaning, and solvent cleaning and oxygen plasma treatment. The sharpest superconducting transition is observed for diamond grown on oxygen plasma treated silicon nitride. Oxidation of a substrate surface under an oxygen plasma should find regular use prior to the growth of CVD diamond on silicon nitride, especially for device applications where control over the nature of the film is critical.

## Methods

1 μm thick, LPCVD non-stoichiometric amorphous silicon nitride thin films on 356 μm p-type boron doped <100> 4" silicon wafers were used as substrates throughout.

**Solvent clean**

The wafers were solvent cleaned prior to any pre-seeding treatment. The wafers were placed into a bath of acetone at 45 °C for 30 min, followed by a methanol bath at room temperature for 5 min, they were then rinsed with deionised water, and dried under nitrogen gas.

**Oxygen plasma treatment**

Oxygen plasma treatment was carried out in a PE-25 Plasma Cleaner using a RF plasma source. The chamber was evacuated at room temperature prior to plasma formation to reduce gaseous contamination. The process was carried out using 30 sccm of oxygen gas flow, a plasma power of 30 W, and a total plasma exposure time of 1 min.

**RCA-1 clean**

The RCA-1 clean was carried out as per the industry standard, using a 5:1:1 ratio of DI $H_2O$ : 30% $H_2O_2$ : 30% $NH_4OH$.[63]

**Zeta Potential**

The ζ potential of silicon nitride was determined from streaming current measurements that were taken using a SurPASS 3 electrokinetic analyser. The SurPASS system comprises two Ag/AgCl electrodes at either end of a streaming channel, and the streaming current is determined by measuring the current flowing towards the outlet electrode, as a function of the electrolyte pressure. The streaming channel is generated by mounting two pieces of silicon nitride (20 mm x 10 mm) face on in parallel, with ~100 μm gap between the wafer faces. An electrolyte flows through the streaming channel shearing counterions from the charged surface of the silicon nitride. The flow of counterions is dependent on the electric double layer created by the surface charge of the silicon nitride. A $10^{-3}$ M solution of potassium chloride was used as the electrolyte solution, and the pressure is controlled between 600 and 200 mbar. The ζ potential is measured as a function of electrolyte pH; 0.1 M HCl and 0.02 M NaOH solutions are used to induce a change in electrolyte pH, using the SurPASS's inbuilt titration system.

Four measurements were taken at each pH value, and an average was taken of the resulting data points.

**X-ray photoelectron spectroscopy**

XPS analysis was carried out using an Al Kα radiation source at 1486.68 eV, operating at 12 kV anode potential and 6 mA emission current. Measurements were taken under an argon environment, to facilitates charge neutralisation. Broad survey scans and narrow scans of the relevant peaks were obtained at pass energies of 150eV and 40 eV respectively. Analysis of data was performed in CasaXPS and peaks were normalised using relative sensitivity factors.

**Substrate seeding**

Prior to film growth, silicon nitride substrates were seeded using a monodisperse aqueous colloid of hydrogen terminated diamond nanoparticles approx. 5 nm in diameter. The process by which diamond particles can be surface terminated with hydrogen can be found elsewhere.[29] The silicon nitride substrate was placed into a seeding solution and ultrasonically agitated for 10 min. The substrate was then rinsed with deionised water, spun dry, and loaded directly into the reaction chamber.

**Diamond film growth**

Diamond films were grown using a Seki Technotron AX6500 series microwave chemical vapour deposition system. The growth temperature was maintained at ~800°C. The substrates were exposed to a gas mixture of methane, hydrogen and trimethylboron, with a 3% $CH_4/H_2$ concentration and a 25000 B/C ratio. 40 Torr pressure and 3.5 kW microwave power were used. After the growth, samples were cooled in hydrogen.

**Scanning electron microscopy**

Scanning electron microscopy (SEM) images of the diamond films were taken using a Raith e-line operating at 20 kV and a 10 mm working distance.

**Resistance measurements**

A Quantum Design physical property measurement system was used to measure each sample's resistance as a function of temperature. Samples were measured in the range 1.9 - 300 K. Silver paste contacts attached to the sample surface using four wires in a van der Pauw configuration. A current of 0.1 µA, 0.1 µA, and 0.5 µA were made to pass through boron-doped diamond grown on solvent cleaned, oxygen plasma treated, and RCA-1 cleaned silicon nitride, respectively. The voltage was measured as the current was passed, and the resistance determined. The resistive transition temperature was taken to be the point at which the resistance diverges by 1% from zero resistance. The resistive transition width is determined to be the temperature range between the resistive transition temperature and the point at which the resistance diverges from the normal state resistance by 1%.

**Raman spectroscopy**

Raman spectroscopy was performed using an inVia Renishaw confocal microscope equipped with a 514 nm laser.

## Data Availability

The datasets generated during and/or analysed during the current study can be found at http://doi.org/10.17035/d.2018.0055370965

## Acknowledgements

This project has been supported by the Engineering and Physical Sciences Research Council under the program Grant GaN-DaME (EP/P00945X/1).

## Author contributions
G.M.K, T.G.J, and O.A.W conceived the idea. H.A.B, E.L.H.T, G.M.K, and S.M conducted the experimental work. H.A.B analysed the results, prepared all figures and tables, and prepared the manuscript. A.P conducted grain size analysis of SEM images. All authors reviewed the manuscript.

## Additional information
**Competing interests**
The authors declare no competing interests.